%% file: sample-acmsmall.tex
\documentclass[sigconf, nonacm]{acmart}

\usepackage{caption}
\usepackage{subcaption}
\usepackage{todonotes}
\usepackage{pifont}
\usepackage{xspace}
\usepackage{CJKutf8}
\usepackage{makecell}
\usepackage{enumitem}
\usepackage{listings}
\usepackage{amsmath}
\usepackage{textcomp}
\usepackage{tabularx}
\usepackage{booktabs}
\usepackage{xcolor} % Required for color

\newcommand{\twovals}[2]{%
  #1\%\,{\footnotesize \textcolor{gray}{(#2\%)}}%
}

\newcommand{\ie}{{\em i.e.,\/ }}
\newcommand{\eg}{{\em e.g.,\/ }}
\newcommand{\vs}{{\em vs.\/ }}

\newcommand{\pb}[1]{\vspace{0.5ex}\noindent{\bf \em #1}\hspace*{.3em}}

\newcommand{\one}{({\em i}\/)\xspace}
\newcommand{\two}{({\em ii}\/)\xspace}
\newcommand{\three}{({\em iii}\/)\xspace}
\newcommand{\four}{({\em iv}\/)\xspace}

\newcolumntype{L}[1]{>{\raggedright\let\newline\\\arraybackslash\hspace{0pt}}m{#1}}
\newcolumntype{C}[1]{>{\centering\let\newline\\\arraybackslash\hspace{0pt}}m{#1}}
\newcolumntype{R}[1]{>{\raggedleft\let\newline\\\arraybackslash\hspace{0pt}}m{#1}}

\begin{document}

\title{Understanding the Consequences of VTuber Reincarnation}

% \affiliation{%

\author{Yiluo Wei}
\affiliation{%
  \institution{The Hong Kong University of Science and Technology (Guangzhou)}
  \country{}}


\author{Gareth Tyson}
\affiliation{%
  \institution{The Hong Kong University of Science and Technology (Guangzhou)}
  \country{}}

% \affiliation{%

% \affiliation{%

% \affiliation{%

% \affiliation{%

% \affiliation{%

% \renewcommand{\shortauthors}{}

\begin{abstract}
% \vspace{-0.5ex}
The rapid proliferation of VTubers---digital avatars controlled and voiced by human actors (Nakanohito)---has created a lucrative and popular entertainment ecosystem. However, the prevailing industry model, where corporations retain ownership of the VTuber persona while the Nakanohito bears the immense pressure of dual-identity management, exposes the Nakanohito to significant vulnerabilities, including burnout, harassment, and precarious labor conditions. When these pressures become untenable, the Nakanohito may terminate their contracts and later debut with a new persona, a process known as ``reincarnation''. This phenomenon, a rising concern in the industry, inflicts substantial losses on the Nakanohito, agencies, and audiences alike. Understanding the quantitative fallout of reincarnation is crucial for mitigating this damage and fostering a more sustainable industry. To address this gap, we conduct the first large-scale empirical study of VTuber reincarnation, analyzing 12 significant cases using a comprehensive dataset of 728K livestream sessions and 4.5B viewer interaction records. Our results suggest reincarnation significantly damages a Nakanohito's career, leading to a decline in audience and financial support, an increase in harassment, and negative repercussions for the wider VTuber industry. Overall, these insights carry immediate implications for mitigating the significant professional and personal costs of the reincarnation, and fostering a healthier and more equitable VTuber ecosystem.

\end{abstract}


% %%
% %% Keywords. The author(s) should pick words that accurately describe
% %% the work being presented. Separate the keywords with commas.
% \keywords{Do, Not, Us, This, Code, Put, the, Correct, Terms, for,


\maketitle

\input{sections/1.Introduction}

\input{sections/2.Background}

\input{sections/3.Method}

\input{sections/4.RQ1}
\input{sections/5.RQ2}

\input{sections/6.RQ3}

\input{sections/7.Conclusion}

\bibliographystyle{ACM-Reference-Format}
\bibliography{sample-base}

\input{sections/A.Appendix}


\end{document}

%% file: sections/1.Introduction.tex
% \vspace{-3ex}
\section{Introduction}

VTubers are digital personas that utilize 2D or 3D animated avatars to engage audiences in livestreams. These avatars are voiced and controlled by a human actor known as the \emph{Nakanohito} (``person inside''), with software such as Live2D \cite{Live2D} to synchronize the actor's movements with the digital character. The phenomenon has experienced explosive growth, resulting in large fanbases and significant corporate sponsorship opportunities \cite{chi25who_reaps, www25virtual}. Currently, there are tens of thousands of active VTubers, the most popular of whom command millions of followers \cite{vtb-fan-ranking, vtb-fan-ranking2}. Financially, the sector is highly lucrative; notably, 31 of the top 50 all-time highest earners via YouTube Superchats (\ie donation) are VTubers, with individual earnings ranging from 1.1 to 3.2 million USD \cite{vtb-superchat-ranking}.

Central to this ecosystem is the Nakanohito. Although early literature suggested viewers might tolerate the replacement of the actor \cite{lu2021kawaii}, subsequent research and real-world examples indicate that the Nakanohito is intrinsic to the VTuber's identity. Thus, replacing the actor results in the character's decline \cite{chi25cant_believe, chi25plave, cscw24_reconstruction}. Consequently, the Nakanohito is an indispensable component: their departure effectively marks the end of the VTuber's existence.

However, despite their critical role, the Nakanohito are a vulnerable population susceptible to various forms of harm and exploitation. A central challenge is the demanding management of a dual identity \cite{lu2021kawaii, Wijaya2023holo}, which often involves a deliberate disembodiment to preserve the avatar's pristine image. This curated performance can require exaggerated emotional or even sexualized expressions to meet audience expectations \cite{cscw24_reconstruction, 10.1145/3604479.3604523}, constituting intensive emotional labor aimed at cultivating parasocial relationships. 

The psychological toll of this continuous output is significant, frequently leading to burnout \cite{cscw24_reconstruction, 10058945, 10.1145/3555104}. Compounding these performance-related pressures are structural issues within the VTuber industry. Corporate ownership of the avatar as intellectual property often leaves the Nakanohito in a precarious labor position with limited control over their virtual persona, creating power imbalances that can lead to disputes \cite{lu2021kawaii, dis24hidden}. Furthermore, the virtual avatar offers only a partial shield against harassment, as the persistent threat of doxing (public exposure of their real-world identity) remains a primary concern \cite{cscw24_reconstruction, Wijaya2023holo, 10.1145/3555104, www25virtual}.

When these situations become untenable, the worst-case scenario unfolds: the Nakanohito terminates their contract with the agency and ceases performing as the current VTuber. 
After some time, the Nakanohito may recover from the issues and may wish to continue their career. 
However, they must debut under a completely new persona since the original VTuber avatar and its associated identity are typically the intellectual property of the corporation.
This process is colloquially known within the community as ``\emph{reincarnation}''.
This phenomenon is largely unique to the VTuber industry and its specific labor dynamics. Unlike traditional streamers whose personal brand and identity are their own, the VTuber's public-facing persona is often a corporate-owned asset. Consequently, the Nakanohito must shed their established character and start anew. Moreover, they are often bound by non-disclosure agreements that prevent them from linking their new identity to their old one, posing challenges for advertising and marketing.

The termination and reincarnation represent a lose-lose-lose situation, inflicting damage on all primary stakeholders. For the Nakanohito, it means abandoning an established identity and community, forcing them to rebuild their career from scratch while losing many of their original fans. For the agency, it results in the loss of a popular and profitable intellectual property. For the viewers, it means the abrupt disappearance of a favorite personality, severing the parasocial bonds they had invested in  \cite{chi25cant_believe, xu2024boundary}.

This vulnerability is inherent to the conventional VTuber model, where the Nakanohito, which is most likely to fail, is a single point of failure of the system. As a result, the industry has witnessed a rising number of abrupt terminations and subsequent reincarnations, including notable cases such as Gawr Gura, the VTuber with the largest following \cite{vtb-fan-ranking, vtb-fan-ranking2}, and Uruha Rushia, the most virtual-gifted VTuber \cite{vtb-superchat-ranking}.
This trend signals an urgent need for a clearer data-driven understanding of these transitions and their repercussions, which has been largely unexplored in prior research.
While such terminations may not always be preventable, insights into their aftermath can help stakeholders better prepare, reduce the associated losses, and inform more resilient and sustainable industry models.
Thus, to address this gap, we provide the first large-scale empirical analysis of the consequences of VTuber reincarnation.

To conduct the analysis, we first compiled a list of 12 significant termination-reincarnation cases, each involving a VTuber who had at least 200K followers (with a maximum of 4.7M and an average of 1.3M) prior to termination. Next, we gathered a list of all VTubers active on YouTube, as indexed by the Virtual YouTuber Wiki \cite{VirtualYouTuberWiki}, in line with methodologies from previous studies \cite{chi25who_reaps, cscw24_reconstruction}. This process yielded a list of 1,972 VTubers. We then collected metadata and interaction records (\eg chat messages, donations) for all the livestream sessions of these VTubers, resulting in a dataset comprising 728,604 livestream sessions and more than 4.5 billion interaction records. This comprehensive dataset enables us to not only capture the audience of the reincarnated VTubers but also understand how audiences move and migrate within the VTuber community on YouTube. 

Using this dataset, we first explore
% 
% 
% \begin{itemize}[leftmargin=*]
    % \item 
    \textbf{RQ1: How does reincarnation impact the Nakanohito's career trajectory in terms of audience size, financial support, and the composition of their new fanbase?}
% \end{itemize}
% 
Our investigation reveals that the Nakanohito loses a substantial portion of their original audience after reincarnation. The behavior of the large amount of ``lost'' viewers has significant implications to the original agency and the wider VTuber industry. This leads us to our second research question:
% 
% \begin{itemize}[leftmargin=*]
    % \item 
    \textbf{RQ2: For the viewers who do not migrate to the reincarnated VTuber, where do they redirect their attention and support?}
% \end{itemize}
% 
Another key concern is harassment. 
If fans can successfully follow the Nakanohito after reincarnation, this may result in harassment also moving to the new instantiation. The contentious nature of a termination and reincarnation could itself generate new forms of harassment from a once-supportive audience. We therefore ask:
% 
% \begin{itemize}[leftmargin=*]
    % \item 
    \textbf{RQ3: How does reincarnation alter the landscape of harassment directed at the VTuber?}
% \end{itemize}
% 
Overall, our findings include:

\begin{itemize}[leftmargin=*]
    \item \textbf{Reincarnation introduces severe career setbacks} (\S\ref{sec:rq1}): After reincarnation, the Nakanohito experiences a drastic reduction in audience size and, most critically, financial support. We find that on average, they lose over 60\% of their original core fanbase and 76\% of their paying supporters. Moreover, they also struggle to cultivate new core supporters but can only dependent on the small, loyal cohort of original fans who successfully migrate. 

    \item \textbf{The ``lost'' audience represents a net loss to the overall industry} (\S\ref{sec:RQ2}): A significant portion of the ``lost'' viewers (including 65\% of the most financially supportive payers) do not migrate to any other VTubers. Instead, they appear to disengage from the ecosystem entirely, indicating that audience loyalty is often tied to a specific VTuber, and its loss results in a net destruction of value for the industry.

    \item \textbf{Reincarnation intensifies, rather than resolves, harassment} (\S\ref{sec:rq3}): Contrary to providing a fresh start, the process is consistently associated with an increase in the proportion of harassing content (a 67.7\% average rise). This toxicity is driven by a dual threat: legacy harassers who successfully track the Nakanohito to their new identity (37.2\%), and a segment of the formerly supportive fanbase that becomes antagonized by the perceived betrayal of the termination (41.4\%).
\end{itemize}

%% file: sections/2.Background.tex
% \vspace{-2ex}
\section{Background \& Related Work}
\label{sec:back}
% \vspace{-0.5ex}
\pb{Primer on VTubers.}
Originating in Japan, VTubers have experienced a surge in popularity since appearing in 2016. While they originally focused on pre-recorded video uploads, the format has evolved, with livestream becoming the dominant medium \cite{10.1145/3604479.3604523}, and YouTube as the primary hub.
A VTuber is a digital avatar animated by a human performer, referred to in Japanese as the ``Nakanohito''. Most VTubers use 2D models powered by software like Live2D \cite{Live2D}. This technology animates the avatar by tracking facial expressions, while desktop commands trigger body gestures. While standard setups are 2D, VTubers with full-body motion capture can use 3D avatars for greater physical expression \cite{chi25vtuber_atelier, 10.1145/3581783.3612094}. Similar to real-person streamers, VTubers can interact with their viewers by reading and responding to chat messages during streams  \cite{www25virtual, lu2021kawaii}.

The VTuber trend began with Kizuna AI's 2016 debut, who coined the term and whose success, including becoming a cultural ambassador for the Japan National Tourism Organization, sparked a widespread movement \cite{Roll_2018}. The COVID-19 pandemic accelerated this growth, as lockdowns increased viewership and pushed VTubing into the mainstream \cite{10.36995, 10.59306}.
The subsequent commercial success, evidenced by VTubers dominating YouTube's Superchat earnings \cite{vtb-superchat-ranking}, led to the establishment of professional talent agencies like Hololive, which have professionalized the industry and driven its global expansion. VTubers are now a major part of online entertainment. Today, there are tens of thousands of active VTubers entertaining global audiences.
The most prominent creators boast millions of followers, earn millions of dollars \cite{vtb-fan-ranking, vtb-superchat-ranking}, and get a substantial  amount of fanart \cite{10.1145/3664647.3680631}.

Previous work on VTubers has examined the motivations driving viewer engagement \cite{www25virtual, lu2021kawaii}, the formation of parasocial relationships for emotional gratification and stress relief \cite{digra2667, tan2023more}, the appeal of their unique aesthetics and creative performances \cite{imx25entertainers, Li2023genz, 10058945}, the distinct community dynamics and behavioral norms \cite{chi25who_reaps, Lill2025ontological}, and the AI-driven VTubers \cite{wei2025even, ye2025favorite}.

\pb{Nakanohito is a Vulnerable Role.}
Research has identified several challenges associated with the Nakanohito model, particularly regarding the strain of managing a dual existence \cite{lu2021kawaii, Wijaya2023holo}. To maintain the avatar's idealized persona, performers often practice a form of disembodiment, which can pressure them into performing exaggerated sexual behaviors to satisfy viewer demands \cite{cscw24_reconstruction, 10.1145/3604479.3604523}. Although these performances reflect aspects of the actor's true self, they require immense emotional labor to sustain parasocial bonds. The psychological cost of this relentless emotional output is high and frequently results in burnout \cite{cscw24_reconstruction, 10058945, 10.1145/3555104}.

Beyond performance pressures, structural inequities within the industry further complicate the Nakanohito’s role. Because corporations typically hold intellectual property rights over the avatars, Nakanohito often lack job security and autonomy over their virtual identities, resulting in significant power disparities \cite{lu2021kawaii, dis24hidden}. Additionally, while an avatar provides a degree of anonymity, it does not strictly shield the Nakanohito from harassment; the threat of doxing and other emerging antisocial behaviors \cite{10.1145/3757690} remains a critical safety concern \cite{cscw24_reconstruction, Wijaya2023holo, 10.1145/3555104, www25virtual}.

\pb{VTuber Termination and Reincarnation.}
Given that the role of the Nakanohito is inherently vulnerable, it is not uncommon for individuals to be unable to continue their activities. A key characteristic of the VTuber model is that the Nakanohito is considered irreplaceable; the persona and the performer are uniquely intertwined \cite{chi25cant_believe, chi25plave, cscw24_reconstruction}. Consequently, the termination of a Nakanohito also results in the permanent termination of the VTuber. It means the avatar and persona are retired and not passed on to another performer. This finality leads many viewers to experience the event as a form of ``death'', causing significant emotional and financial loss for the performer, the managing agency, and the audience alike~\cite{chi25cant_believe}. 

The reasons for termination often stem from the role's inherent vulnerabilities, including inappropriate behavior, mental health struggles, harassment, or conflicts with the agency. Since many of these issues are situational, a performer may resolve them after detaching from their original VTuber identity. It is therefore common for a former Nakanohito to return to streaming with a new avatar and persona, a phenomenon the community refers to as ``reincarnation''. 
The reincarnated VTuber usually would not officially announce or declare any connection to the terminated identity (likely due to contractual agreements with the agency). By community convention, viewers also refrain from mentioning this connection in more formal situations, although often times it has already become common knowledge.
This practice creates a complex perceptual problem, as viewers' interpretations of the link between the original and reincarnated VTuber vary widely, from seeing them as the same entity to a mere substitute, or as completely different individuals, often causing conflict within the community. Previous research has also examined the complex relationship and varied viewer perceptions regarding the connection between the Nakanohito and the virtual persona  \cite{Lill2025ontological, tang2025broadcast, cscw24_reconstruction, Xu_Niu_2023, Turner1676326, chi25plave}. 

Existing research on reincarnation is sparse and largely qualitative. The most relevant study is \cite{chi25cant_believe}, which qualitatively analyzes parasocial grieving in response to VTuber termination. Other works have also qualitatively touched upon viewer perceptions in this context \cite{10.1177/14614448231212822, sasabe2024vtuber, xu2024boundary}. We provide the first large-scale, quantitative analysis of into this complex matter of VTuber reincarnation.

%% file: sections/3.Method.tex
% \vspace{-2ex}
\section{Data Collection \& Processing}
% \vspace{-0.5ex}
\label{sec:method}

\pb{Selecting Reincarnation Cases.}
The initial step in our study is to compile a list of VTuber terminations and reincarnations. Our goal is to identify significant and representative cases. To achieve this, we begin by ranking VTubers according to their number of followers \cite{vtb-fan-ranking2}. Next, we filter this list to include only those VTubers who have both terminated and reincarnated. 
We set a threshold of 200,000 followers, which approximately corresponds to the top 500 most popular VTubers. 
This approach ensures we capture cases with significant impact. Furthermore, having a large follower count means these VTubers receive substantial attention within the community. As a result, while reincarnations cannot be officially announced, they are widely acknowledged within VTuber fan communities.

These fan communities include Reddit's \texttt{r/VirtualYoutubers}, a major English-language forum for VTuber discussions, which has also been utilized in previous research \cite{chi25cant_believe}; the NGA VTuber forum \cite{nga}, a major Chinese platform for VTubers that has also been referenced in prior research \cite{lu2021kawaii}; and 5ch streaming \cite{5ch}, a major Japanese forum for livestream and VTubers. 
Additionally, reincarnations are often validated through subtle hints on the Nakanohito's personal social media accounts, as well as clues shared by the terminated VTuber's friends on their accounts.
Thus, we review these resources to identify cases where we can be extremely confident that the reincarnation is authentic.

Overall, this process results in a list of 12 VTubers who have undergone reincarnation. Table \ref{tab:vtb_reincarnation} provides a summary. In accordance with the conventions of the VTuber industry and community, we have anonymized the VTubers in the table to avoid explicitly linking the original VTuber and reincarnated VTuber.

\begin{table}[h!]
    \centering
    \small
    \resizebox{0.95\linewidth}{!}{
    \begin{tabular}{|rrrr|r|}
        \toprule
        \textbf{Original} & \textbf{Affiliation} & \textbf{Termination} & \textbf{\# Followers} & \textbf{Reincar.} \\
        \midrule
        V1  & Nijisanji & 2024-01-20 & 579K & R1 \\
        V2  & Nijisanji & 2024-06-12 & 552K & R2 \\
        V3  & Nijisanji & 2023-07-08 & 538K & R3 \\
        V4  & Nijisanji & 2022-04-30 & 220K & R4 \\
        V5  & Hololive  & 2021-07-01 & 1.37M & R5 \\
        V6  & Hololive  & 2024-08-28 & 2.24M & R6 \\
        V7  & Hololive  & 2025-01-03 & 1M    & R7 \\
        V8  & Hololive  & 2024-10-01 & 1.77M & R8 \\
        V9  & Hololive  & 2025-01-26 & 1.35M & R9 \\
        V10 & Hololive  & 2025-05-01 & 4.72M & R10 \\
        V11 & Hololive  & 2025-04-26 & 1.34M & R11 \\
        V12 & Other     & 2023-09-01 & 210K  & R12 \\
        \bottomrule
    \end{tabular}}
    \caption{Information of the VTubers included in this study.}
    \vspace{-6ex}
    \label{tab:vtb_reincarnation}
\end{table}

\pb{VTubers \& Their Basic Information.}
To study audience retention and migration, we need data not only on reincarnated cases but also, ideally, on all VTubers on YouTube. While there is no comprehensive source or index for all VTubers, we follow previous research \cite{chi25who_reaps, cscw24_reconstruction} and make use of the Virtual YouTuber Wiki \cite{VirtualYouTuberWiki}, which catalogs most of the popular VTubers along with their information. In total, we compile a dataset of 1,972 VTubers who stream on YouTube, including their basic information (\eg affiliation).

\pb{Livestream Sessions \& Interaction Records.}
For all the VTubers in our dataset, we gather data from their livestream sessions on YouTube. We first collect the metadata for each livestream, such as title and start time; please refer to Table \ref{tab:live_description} for the detailed structure of this meta information. Following this, we gather all interaction records from the sessions, covering four features: \one real-time chat messages, which allow viewers to communicate with the streamer and each other; \two Superchats (SC), used for direct financial support and increased visibility via highlighted, pinned messages; \three Memberships, which represent sustained support through recurring monthly subscriptions (typically 4.99 USD); and \four Gift Memberships, a mechanism where users purchase memberships that are randomly distributed to other viewers to foster community support. The structure of these interaction records is detailed in Table \ref{tab:interaction_description}. In total, we compile a dataset consisting of 728,604 livestream sessions and 4,552,865,327 interaction records.

\pb{Defining Time Period.}
To analyze the impact of termination and reincarnation, we establish two  observation periods. 
The \emph{before termination} period covers the 30-day interval prior to the VTuber's final livestream. We selected an one-month window because it corresponds to the standard duration of channel membership subscriptions. However, to guarantee a sufficient data sample size, we require a minimum of 20 livestream sessions; this threshold was chosen to approximate the average monthly streaming frequency for popular VTubers. Consequently, if the 30-day interval contains fewer than 20 sessions, the period is extended to encompass the last 20 livestreams.
The \emph{after reincarnation} period is defined symmetrically, starting from the reincarnated VTuber's first livestream. This period covers the first 30 days, or is extended to the first 20 livestreams if that threshold is not met.

\pb{Coding the Viewers.}
In our analysis, we categorize viewers into three groups:
A viewer is classified as a \emph{Payer} if they have made any financial contributions (superchats, memberships, or gift memberships) to the VTuber during the specified time period.
A viewer is an \emph{Active Viewer} if they are not a payer and have interaction records in at least two livestream sessions of the VTuber within the time period.
All other viewers are thus labeled as \emph{Inactive Viewers}.
This classification is meant to differentiate between random viewers who occasionally watch a livestream session and regular viewers.

\pb{Identifying Harassment Chat Messages.}
In our analysis, we need to identify hate speech and harassment messages directed at VTubers. We utilize the OpenAI Moderation API \cite{moderationAPI}, employing the \texttt{omni-moderation} model \cite{moderationModel}. This model is capable of detecting various types of multilingual toxic content. Among them, the ``harassment'' type, defined as ``\textit{Content that expresses, incites, or promotes harassing language towards any target}'', aligns most closely with our objective. Therefore, for each chat message, we call the OpenAI Moderation API, and if the result flags the message as harassment, we label it accordingly.

%% file: sections/4.RQ1.tex
\vspace{-1ex}
\section{RQ1: The Losses of the VTubers}
\label{sec:rq1}

In this section, we explore RQ1: How does reincarnation impact the career trajectory of the Nakanohito? It is anticipated that they may lose viewers, fans, and financial support.

\vspace{-1ex}
\subsection{Overall Livestream Popularity}
\begin{figure}[h!]
    \centering
    % \vspace{-2ex}
    \includegraphics[width=\linewidth]{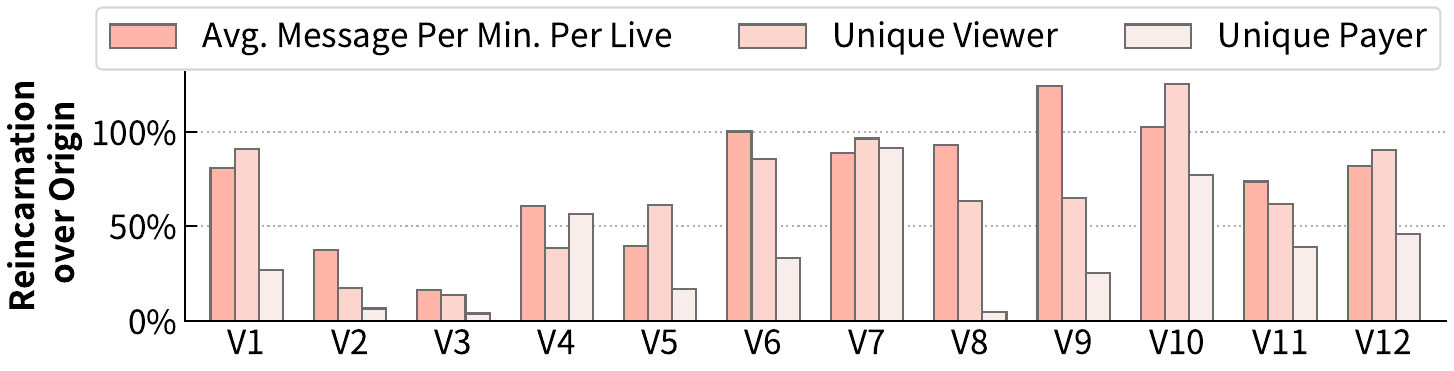}
    \vspace{-5ex}
    \caption{Overall popularity difference of the livestreams before termination and after reincarnation.}
    \vspace{-2ex}
    \label{fig:overall_live_popularity}
\end{figure}

We begin by comparing livestream popularity before the termination of the original VTuber against those after the reincarnation. This analysis provides an overview of the losses in audience engagement and support.
To quantify popularity, we employed three metrics: \one the average number of chat messages, \two the number of unique viewers, and \three the number of unique payers. We calculate the ratio of post-reincarnation to pre-termination values for each metric. The results are presented in Figure \ref{fig:overall_live_popularity}.

The figure reveals a trend of decline across all metrics. For the number of chat messages, a general decrease is observed (average -33.4\%), though there are exceptions like V9 and V10 which slightly surpass their original levels.
A similar pattern is evident for the number of unique viewers, where nearly all individuals experience a decline (average -42.8\%). While the previous two metrics indicate general audience engagement, which can fluctuate, the most significant and consistent impact is observed in the number of payers, who represent the core fan base. Every case experiences a significant drop in unique payers, a metric that consistently shows the lowest result (average -76.3\%).
Overall, the results confirm that the reincarnation results in a decline in the popularity of VTubers, which is particularly severe regarding financial support. 

\vspace{-2ex}
\subsection{Viewers' Migration Directions}
Next, to better understand the dynamics behind the decline in popularity, we investigate the migration patterns of the audience. 

\begin{figure}[h!]
    \centering
    % \vspace{-2ex}
    \includegraphics[width=\linewidth]{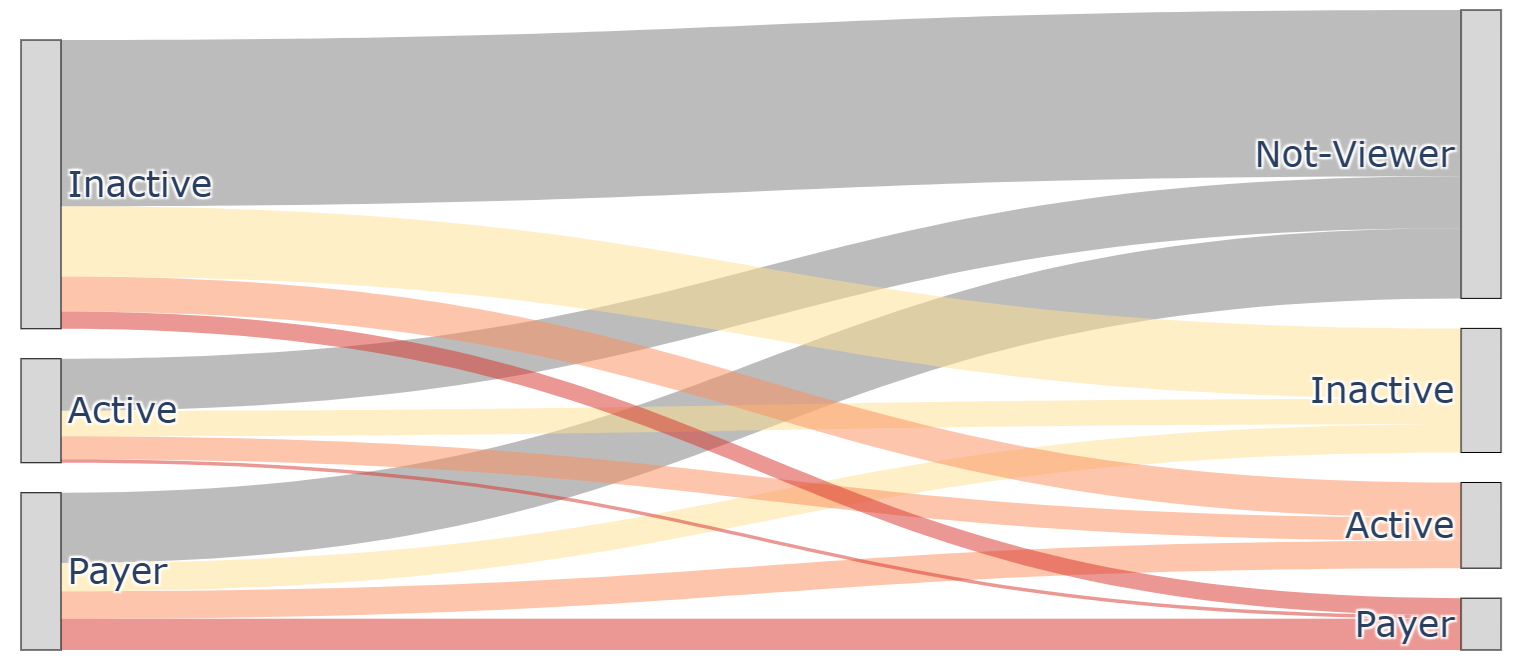}
    \resizebox{0.99\linewidth}{!}{
    \begin{tabular}{|r|r|rrrr|}
    \toprule
        \textbf{Original} & \textbf{Total} & \textbf{Payer} & \textbf{Active} & \textbf{Inactive} & \textbf{Non-Viewer} \\
    \midrule
        \textbf{Payer} & 8.82\% & \twovals{13.13}{1.16} & \twovals{12.45}{1.10} & \twovals{13.23}{1.17} & \twovals{61.19}{5.40} \\
        \textbf{Active} & 4.83\% & \twovals{1.83}{0.09} & \twovals{17.57}{0.85} & \twovals{20.04}{0.97} & \twovals{60.55}{2.93} \\
        \textbf{Inactive} & 86.34\% & \twovals{0.67}{0.58} & \twovals{1.76}{1.52} & \twovals{6.23}{5.38} & \twovals{91.34}{78.86} \\
    \bottomrule
    \end{tabular}}
    \vspace{-2ex}
    \caption{Left: Original viewers of the VTuber before termination; Right: What original viewers become for the the reincarnated VTuber. The flow size is log-scaled for visualization. Table: numeric value of the flow.}
    \vspace{-2ex}
    \label{fig:original_viewer_go_sankey}
\end{figure}

\pb{Where does the Original Viewer Go?} 
We first examine the perspective of the original viewers. That is, where do the viewers of the original VTuber go after reincarnation? Figure \ref{fig:original_viewer_go_sankey} presents a Sankey diagram illustrating the average migration flow across all 12 cases, with the table providing the numerical values corresponding to the flows depicted in the diagram.

The data reveal a significant loss of the most engaged audience segments. Of the original Payer group, which constituted 8.82\% of the initial audience, only 13.13\% continue as payers after the VTuber's reincarnation. The majority of these core supporters either become inactive (13.23\%) or, most commonly, non-viewers (61.19\%), indicating they do not migrate at all. 
A similar trend holds for active viewers, of whom only 17.57\% retain their active status, while the majority (60.55\%) also become non-viewers.
Conversely, the reincarnation process does reactivate a small portion of the previously inactive audience. Of this large group, 0.67\% transition to becoming payers and 1.76\% become active after reincarnation. While this demonstrates an opportunity to re-engage a dormant audience, this gain is numerically small compared to the loss of active viewers and payers.

Overall, the results highlight that a VTuber loses a substantial portion of their most loyal fans and financial supporters after reincarnation. This failure to retain the core fanbase presents a significant challenge, as previous research suggests that a dedicated community is a key factor in a VTuber's long-term success and sustainability \cite{www25virtual, lu2021kawaii, chi25who_reaps}.
However, there remains the possibility of successfully attracting new core fans who were not part of the original audience. We explore this in the following paragraph.

\begin{figure}[h!]
    \centering
    \vspace{-2ex}
    \includegraphics[width=\linewidth]{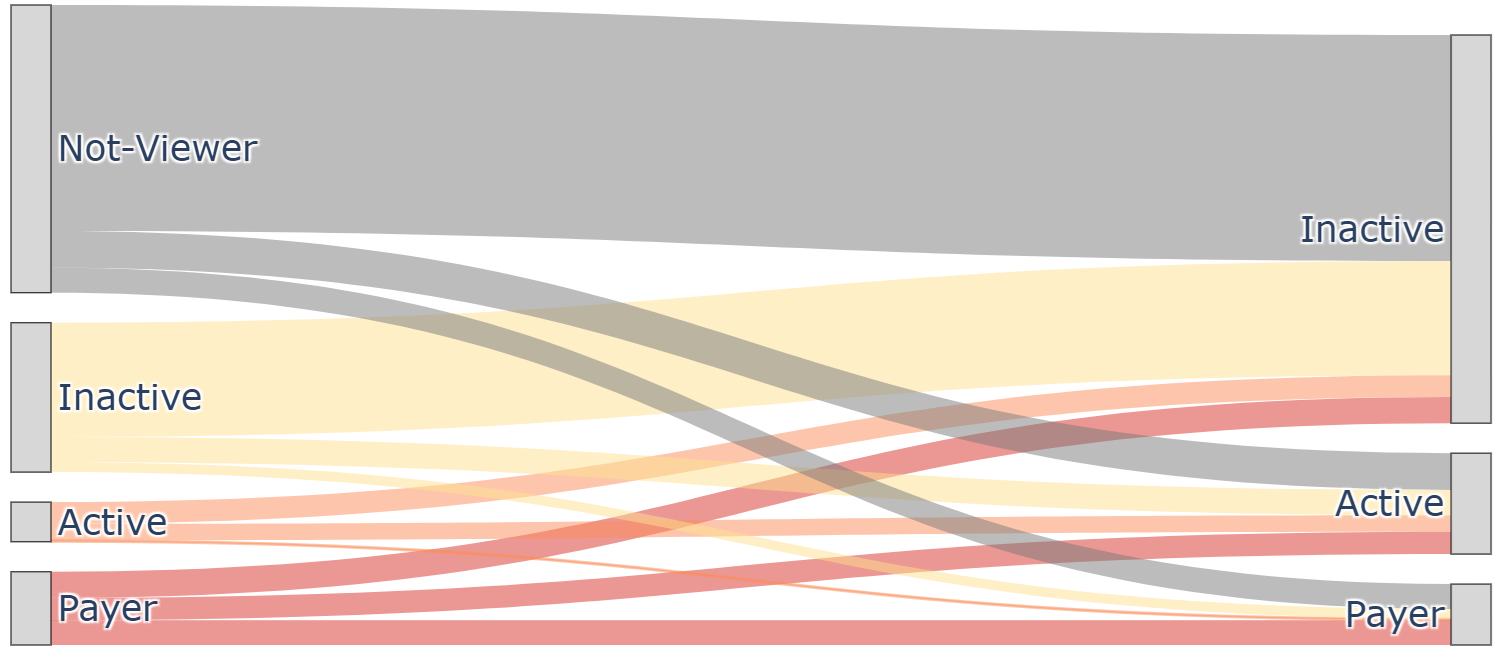}
    \resizebox{0.995\linewidth}{!}{
    \begin{tabular}{|r|r|rrrr|}
    \addlinespace
    \toprule
        \textbf{Reincar.} & \textbf{Total} & \textbf{Payer} & \textbf{Active} & \textbf{Inactive} & \textbf{Non-Viewer} \\
    \midrule
        \textbf{Payer} & 11.07\% & \twovals{40.69}{4.51} & \twovals{2.39}{0.27} & \twovals{15.98}{1.77} & \twovals{40.94}{4.53} \\
        \textbf{Active} & 18.35\% & \twovals{22.20}{4.07} & \twovals{16.22}{2.98} & \twovals{25.18}{4.62} & \twovals{36.41}{6.68} \\
        \textbf{Inactive} & 70.58\% & \twovals{6.71}{4.73} & \twovals{5.63}{3.97} & \twovals{29.41}{20.76} & \twovals{58.25}{41.11} \\
    \bottomrule
    \end{tabular}}
    \vspace{-2ex}
    \caption{Right: Viewers of the reincarnated VTuber; Left: What these viewers were for the original VTuber. Table: numeric value of the flow.}
    \vspace{-2ex}
    \label{fig:new_viewr_come_sankey}
\end{figure}

\pb{Where New Viewers Come From?}
Next, we examine the perspective of the viewers of the reincarnated VTuber. That is, where do the viewers of the reincarnated VTuber come from? Figure \ref{fig:new_viewr_come_sankey} presents the Sankey diagram illustrating the average migration flow across all 12 cases, with the table providing the numerical values corresponding to the flows depicted in the diagram.

The results show that while reincarnated VTubers successfully attract a large cohort of new viewers (those who were non-viewers of the original channel), these newcomers primarily contribute to inactive viewership. The majority of these new arrivals (78.57\%) are Inactive viewers in the new channel, suggesting they are not part of the core fanbase.
In stark contrast, the new core fanbase, comprising payers and active viewers, is predominantly formed by the audience from the original channel. Specifically, over 59\% of the new Payers and over 63\% of the new active viewers were the viewers of the original VTuber. This reveals a disappointing situation: although the previous analysis showed that only a small proportion of original core fans migrate, this small, retained group constitutes the majority of the reincarnated VTuber's core fans.

Collectively, the two perspectives highlight a significant challenge for a reincarnated VTuber. They not only lose a large portion of their original dedicated fanbase but also struggle to cultivate new core supporters from the influx of new viewers. This dual difficulty in both retaining old fans and converting new ones poses a substantial obstacle to rebuilding a sustainable, engaged community.

\begin{figure}[h!]
    \centering
    \includegraphics[width=\linewidth]{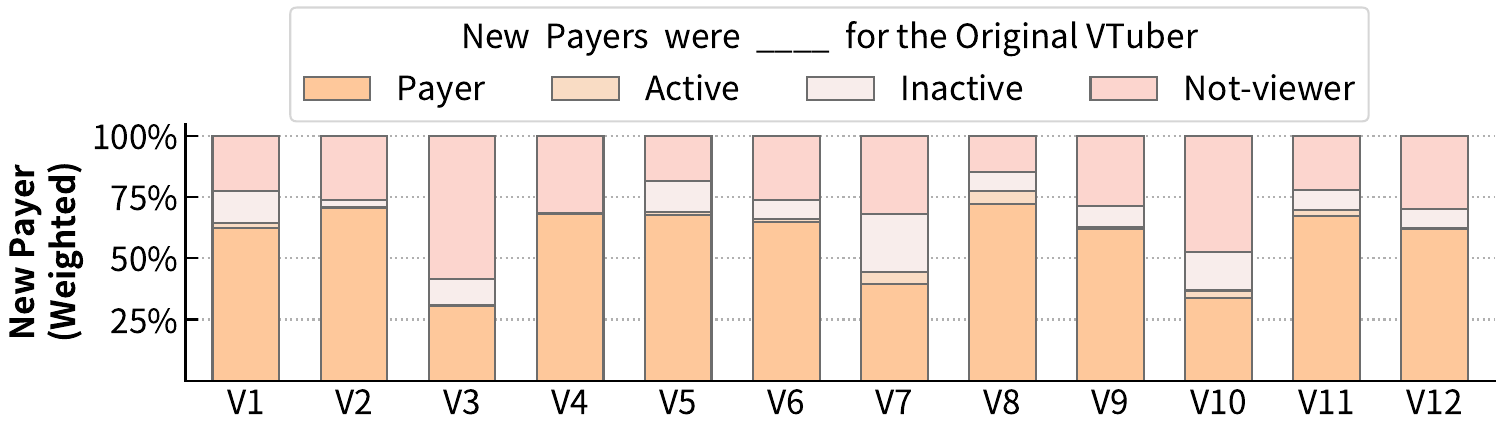}
    \vspace{-5.5ex}
    \caption{The weighted composition of the payers.}
    \vspace{-3ex}
    \label{fig:new_payer_weighted_come_bar}
\end{figure}

\pb{Where the Financial Support Comes From?}
The previous analysis only concern the payer counts. To directly assess the financial impact of reincarnation, we analyze the flow of income. We created a migration graph with payers weighted by their monetary contributions. Figure \ref{fig:new_payer_weighted_come_bar} illustrates the results, indicating that ``$X\%$ of the reincarnated VTuber's income originates from original payers, active viewers, inactive viewers, and non-viewers.''

The data reveals that the majority of the reincarnated VTuber's monetary income comes from original Payers. On average, they constituent 41\% of the payers (as shown in the previous analysis), but contribute 58.5\% of the financial support. 
In contrast, new viewers of the reincarnated VTuber (\ie Non-viewers of the original VTuber) also constitute 41\% of the payers, but contribute a significantly smaller share of the income (average 29.6\%). 

This analysis reinforces the findings of the previous analysis, suggesting that the financial viability of a reincarnated VTuber is more heavily dependent on the small, loyal cohort of core fans who migrate with them. Nevertheless, this reliance is not necessarily a negative thing. Their situation is still much better than that of a typical new VTuber, who must strive hard to survive (most VTubers cease operations within three years \cite{chi25who_reaps}). Overall, considering the results in this subsection, reincarnated VTubers may benefit from prioritizing the satisfaction of original fans, rather than competing with all the new VTubers for viewers.

\vspace{-1.5ex}
\subsection{When do the Original Viewers Migrate?}
Given that a reincarnated VTuber cannot explicitly advertise their new identity due to contractual or professional reasons, communication with the original audience is inherently difficult. We therefore hypothesize that it will take time for the information to disseminate and for viewers to discover and migrate to the new channel. To investigate this, we analyze in which livestream session after the reincarnation do the viewer successfully migrate.

\begin{figure}[h!]
    \centering
    \vspace{-2ex}
    \includegraphics[width=0.85\linewidth]{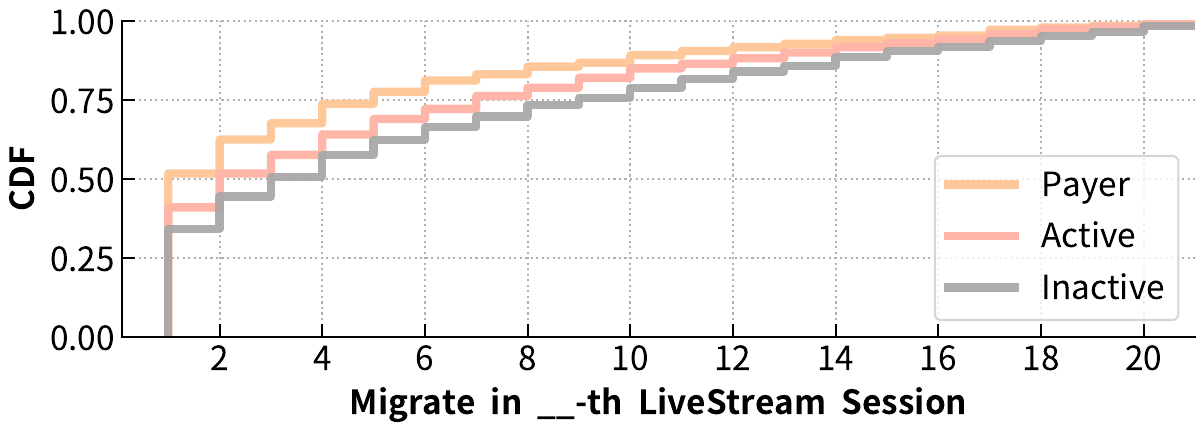}
    \vspace{-3ex}
    \caption{CDF of the order of the livestream sessions in which the viewer successfully migrate after reincarnation.}
    \vspace{-3ex}
    \label{fig:migrate_in_nth_live_cdf}
\end{figure}

Figure \ref{fig:migrate_in_nth_live_cdf} presents the result in a CDF. The plot reveals that the migration is a slow process across all segments. In the debut livestream, only about 51\% of the payers, 41\% of active viewers, and 33\% of inactive viewers from the original audience manage to migrate. This discovery process remains gradual; even after 10 livestream sessions, a significant fraction of the eventual migrating audience (14\% of payers and 17\% of active viewers) has still not arrived.

These findings highlight a significant information dissemination challenge. It suggests one potential explanation for why many original fans do not migrate: they may fail to do so simply because they are unaware of the reincarnation, or they discover it too late, after the initial momentum has passed. That is, a portion of the audience loss might not be a rejection of the new persona, but a direct consequence of the information gap during reincarnation. This highlights the need for improved post-reincarnation discovery mechanisms to facilitate fan migration as effectively as possible.

%% file: sections/5.RQ2.tex
\vspace{-1.5ex}
\section{RQ2: Impact to the VTuber Industry}
\label{sec:RQ2}

Our findings for RQ1 show that reincarnation leads to significant audience attrition, with 61\% of the original payers and active viewers failing to migrate. The destination of this high-value ``lost'' viewers is a critical question for the VTuber industry, especially when the original VTuber had millions of followers. Accordingly, this section investigates where they go after the original VTuber's termination.

\pb{Definitions.}
First, we define ``lost'' viewers: active viewers and payers of the original VTuber $V_0$, who do not continue as viewers of the reincarnated VTuber $R_0$.
Then, we determine where they migrate by comparing their activity before the termination of $V_0$ \vs after the reincarnation of $R_0$.
We consider they migrate to another VTuber $V$, if they were inactive or a non-viewer for $V$ before $V_0$'s termination, and become active for $V$ after the $R_0$'s reincarnation.

% \vspace{-1ex}
\subsection{Migrate to the Same \vs Other Agency}
\label{subsec:migrate_agency}
\begin{figure}[h!]
    \centering
    % \vspace{-2ex}
    \includegraphics[width=\linewidth]{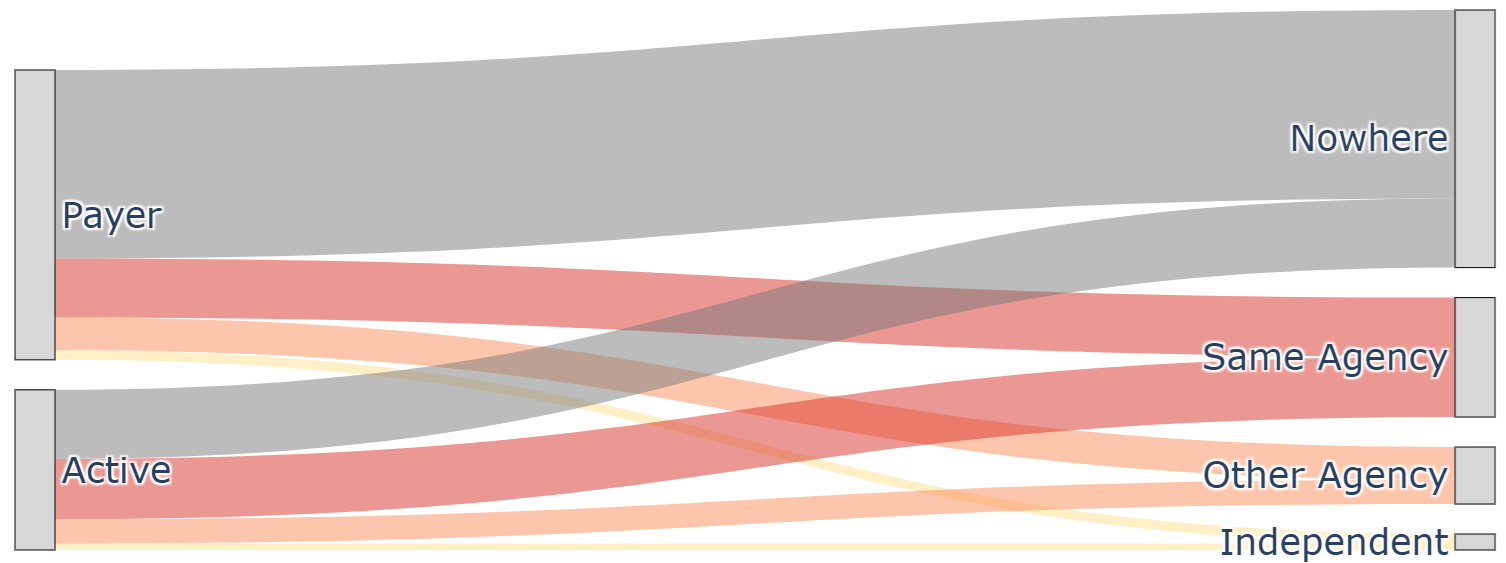}
    \resizebox{0.99\linewidth}{!}{
    \begin{tabular}{|r|r|rrrr|}
    \addlinespace
    \toprule
    \textbf{Lost} & {} & \multicolumn{4}{c|}{\textbf{The Lost Viewers Migrate to VTubers}} \\
    \textbf{Viewers} & \textbf{Total} & \textbf{Same Agency} & \textbf{Other Agency} & \textbf{Independent} & \textbf{Nowhere} \\
    \midrule
    \textbf{Active} & 35.60\% & \twovals{37.64}{13.40} & \twovals{15.13}{5.39} & \twovals{4.05}{1.44} & \twovals{43.18}{15.37} \\
    \textbf{Payer} & 64.40\% & \twovals{20.40}{13.14} & \twovals{11.30}{7.28} & \twovals{3.27}{2.11} & \twovals{65.02}{41.88} \\
    \bottomrule
    \end{tabular}}
    \vspace{-2ex}
    \caption{Left: those who were the viewer of the original VTuber and do not migrate to the reincarnated VTuber; Right: where (which other VTubers) they migrate to, categorized by agency. Table: numeric value of the flow.}
    \vspace{-2ex}
    \label{fig:lost_viewer_go_agency_sankey}
\end{figure}

We first investigate where the ``lost'' viewers go from the agency's perspective.
Figure \ref{fig:lost_viewer_go_agency_sankey} presents the Sankey diagram of the average migration flow across all 12 cases, with the table providing the numerical values corresponding to the flows depicted in the diagram.

We see that a substantial number (65.02\%) of lost Payers fall into the Nowhere category, meaning they do not transfer their support to any other VTuber in our dataset. A similar, albeit less severe, trend is observed for Active viewers, with 43.18\% also not going to any other VTubers in the ecosystem. This may indicate a significant loss for the VTuber industry, and we further explore this in \S\ref{subsec:rq2_nowhere}.

The agency is moderately successful at retaining general engagement, as 37.64\% of the lost Active viewers and 20.4\% of the lost payers redirect their attention to other talents within the same company. 
Beyond the original agency, a smaller yet notable portion of the lost audience migrates to competitors, with 15.13\% active viewers and 11.3\% payers moving to other agencies. They also show a moderate tendency to explore, with 4.05\% active viewers and 3.27\% payers migrating to independent VTubers.

Overall, these findings present a complex picture with strategic implications for all parties. Although the original agency suffers significant losses, it still captures the largest share of the audience that is redistributed within the ecosystem. This dynamic suggests that losses could be better mitigated through direct corporate action and through the efforts of other VTubers within the same roster who can strategically engage and guide displaced fans through collaborations and community events. For other agencies and independent VTubers, a major talent's departure represents a ``whale fall'', \ie a rare opportunity to attract a highly engaged, unattached audience if they can successfully appeal to these displaced fans.

% \vspace{-2ex}
\subsection{Migration to Familiar \vs New VTubers}
\begin{figure}[h!]
    \centering
    \vspace{-2ex}
    \includegraphics[width=\linewidth]{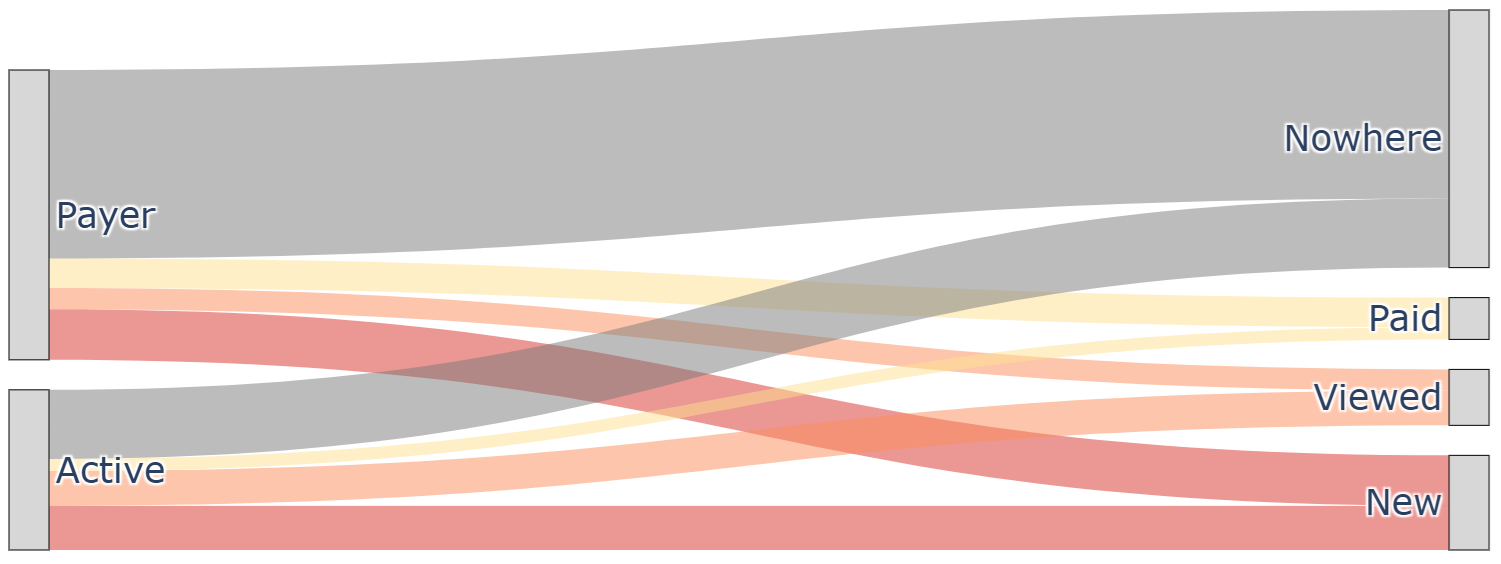}
    \resizebox{0.99\linewidth}{!}{
    \begin{tabular}{|r|r|rrrr|}
    \toprule
    \textbf{Lost} & {} & \multicolumn{4}{c|}{\textbf{The Lost Viewers Migrate to VTubers}} \\
    \textbf{Viewers} & \textbf{Total} & \textbf{Viewed} & \textbf{Paid} & \textbf{New} & \textbf{Nowhere} \\
    \midrule
    Active & 35.60\% & \twovals{21.71}{7.73} & \twovals{7.56}{2.69} & \twovals{27.54}{9.80} & \twovals{43.18}{15.37} \\
    Payer & 64.40\% & \twovals{7.30}{4.70} & \twovals{10.24}{6.59} & \twovals{17.43}{11.23} & \twovals{65.02}{41.88} \\
    \bottomrule
    \end{tabular}}
    \vspace{-2ex}
    \caption{Left: those who were the viewer of the original VTuber and do not migrate to the reincarnated VTuber; Right: where (which other VTubers) they migrate to, categorized by familiarity to the viewer. Table: numeric value of the flow.}
    \vspace{-2ex}
    \label{fig:lost_viewer_go_history_sankey}
\end{figure}

To further dissect the behavior of the non-migrating audience, we next examine whether these viewers gravitate towards VTubers they were already familiar with or if they use the opportunity to discover new ones. We categorize their destinations as VTubers who they had previously \emph{Viewed}, previously \emph{Paid}, or are entirely \emph{New} to them.
Figure \ref{fig:lost_viewer_go_history_sankey} presents the Sankey diagram of the average migration flow across all 12 cases, with the table  providing the numerical values corresponding to the flows in the diagram.

Surprisingly, the data reveals a near-even split between two migration destinations. For lost active viewers, 27.54\% migrate to New VTubers, a figure comparable to the combined 29.27\% who move to VTubers they had already viewed or paid before. This split is also pronounced among lost payers, with 17.43\% moving to New VTubers and an almost identical 17.54\% shifting to familiar ones.

Overall, the result reveals a dual opportunity for the rest of the VTuber ecosystem, confirming the ``whale fall'' phenomenon discussed previously. First, it shows that other VTubers can successfully re-engage their own dormant viewers, \ie those who were previously focused on the terminated VTuber. Second, and more importantly, the reincarnation of a major VTuber creates a rare opening to attract a proven, high-value audience that is now actively seeking a new community to join. These findings suggest a clear strategy for other VTubers: during such industry shifts, they should actively increase their visibility and foster a welcoming environment to effectively capture these migrating high-value fans.

% \vspace{-2ex}
\subsection{Migrate to Nowhere?}
\label{subsec:rq2_nowhere}

\begin{figure}[h!]
    \centering
    \vspace{-2ex}
    \includegraphics[width=\linewidth]{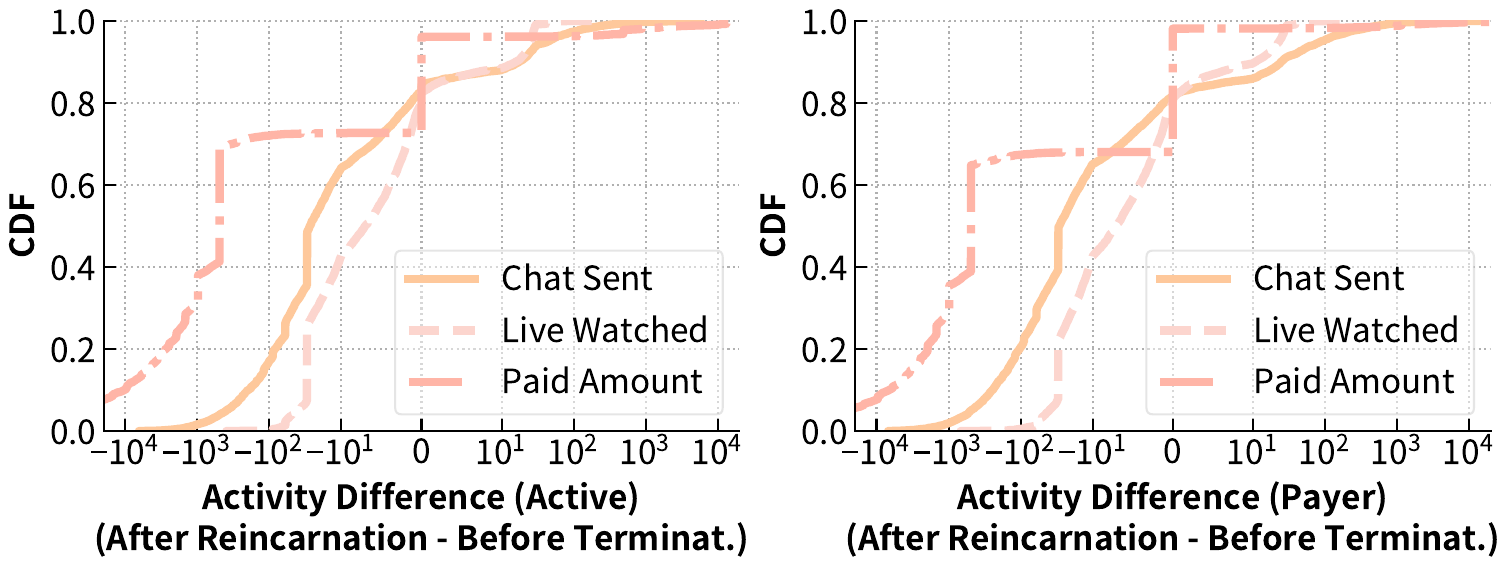}
    \vspace{-4.5ex}
    \caption{CDF of the activity difference for those who were the viewers of the original VTuber and do not migrate to anywhere after reincarnation. (a) for the original active viewers; (b) for the original payers.}
    \vspace{-2ex}
    \label{fig:lost_viewer_nowhere_activity_diff}
\end{figure}

Finally, we investigate the large contingent of viewers who migrate to nowhere. As established previously (\S\ref{subsec:migrate_agency}), this group constitutes the majority of the lost viewers (43.18\% of Active viewers and 65.02\% of Payers). However, this does not necessarily mean they disappeared completely. 
For instance, a viewer might have followed multiple VTubers; when one terminates, they may simply continue their existing viewing habits with others without making a new ``migration''.
To further explore this, we analyze the overall activity change for these viewers by comparing their behavior before termination and after reincarnation. Figure \ref{fig:lost_viewer_nowhere_activity_diff} displays the CDF for the difference in three metrics: \one number of chat messages sent, \two number of livestream sessions watched, and \three amount paid. 

The results show a precipitous drop in engagement across the board. Financial contributions are the most severely affected, where virtually all of these viewers decrease their spending, with the vast majority ceasing financial support entirely. A similar, though less absolute, decline is evident in the number of chat messages sent, average -81.3 (-61.1\%) for active viewer and -84.0 (-42.7\%) for payer; and the livestream watched, average -11.1 (-59.3\%) for active viewer and -11.0 (-49.9\%) for payer.

Overall, the result confirms that the viewers who migrate to nowhere are disappearing: they drastically reduce or eliminate their engagement particularly regarding financial support. This finding demonstrates that a large portion of viewers' loyalty and financial support is specifically tied to individual VTubers. Consequently, when a VTuber is terminated/reincarnated, it leads to a direct and substantial loss of community engagement and revenue for the VTuber industry as a whole. Agencies should therefore re-evaluate their strategies regarding talent retention, as the reincarnation not only risks individual revenue streams but also contributes to a net loss for the entire industry, making the ecosystem more fragile.

%% file: sections/6.RQ3.tex
\vspace{-2ex}
\section{RQ3: Transference \& Evolution of Harassment}
\label{sec:rq3}

Previous sections highlight the substantial audience and financial losses associated with reincarnation. However, another crucial consequence is the evolution of harassment. If fans are able to successfully track the Nakanohito after reincarnation, it is reasonable to assume that some determined harassers can do so as well. Moreover, the controversial nature of termination and reincarnation could potentially provoke new forms of antagonism from an audience that was once supportive. This section explores this further.

\vspace{-1ex}
\subsection{Increase of Harassment}
\begin{figure}[h!]
    \centering
    % \vspace{-2ex}
    \includegraphics[width=\linewidth]{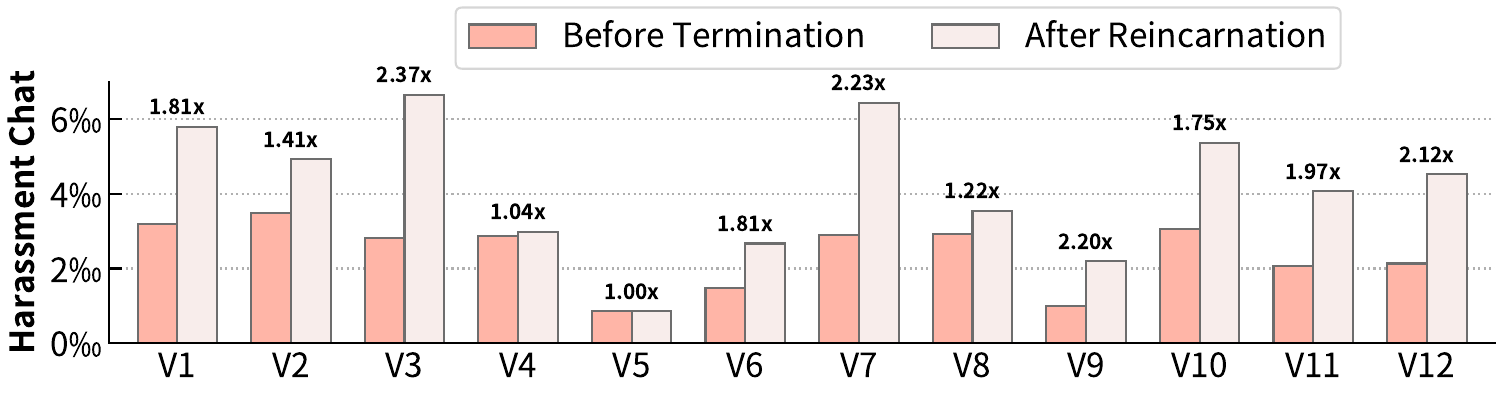}
    \vspace{-5ex}
    \caption{The proportion of harassment chat messages before termination and after reincarnation.}
    \vspace{-3ex}
    \label{fig:harassment_chat_proportion}
\end{figure}

\vspace{-0.5ex}
We begin our analysis by quantitatively assessing the overall change in the proportion of harassment chat messages to provide an overview of the issue. 
To do this, we use the OpenAI moderation API to determine whether a chat message is harassment as described in \S\ref{sec:method}. 
The result is presented in Figure \ref{fig:harassment_chat_proportion}, 
We observe a significant increase in the proportion of harassing messages following before termination \vs after reincarnation.
With the sole exception of V5, where the proportion remained almost unchanged, the other 11 VTubers in our sample experienced a relative increase in harassment. The magnitude of this increase is often substantial (average 67.7\%). For instance, V3 experienced the largest relative rise, with the proportion of harassing chat increasing by a factor of 2.37x. Several others, including V7 (2.23x), V9 (2.20x), and V12 (2.12x), saw the share of harassment in their chats more than double.

These findings confirm that the reincarnation process does not offer a reprieve from harassment. Instead, the transition is consistently associated with a chat environment that becomes proportionally more toxic and hostile for the VTuber.

\vspace{-1.5ex}
\subsection{Composition of the Harassers}
\begin{figure}[h!]
    \centering
    \vspace{-2ex}
    \includegraphics[width=\linewidth]{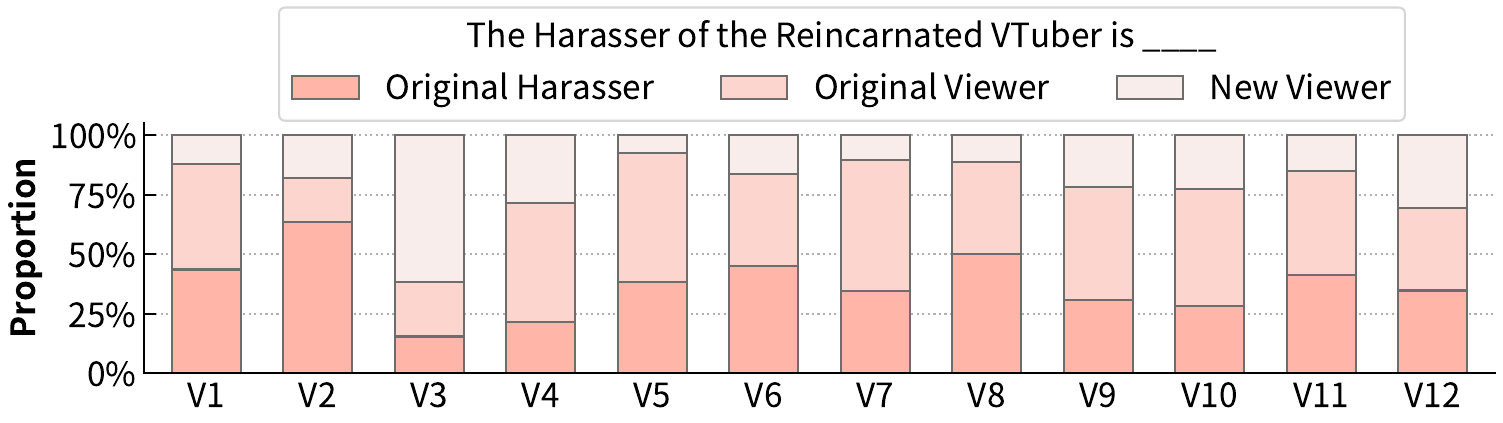}
    \vspace{-5.5ex}
    \caption{The composition of the harassers for the VTuber.}
    \vspace{-2ex}
    \label{fig:harassment_viewer_composition}
\end{figure}

\vspace{-0.5ex}
To understand the source of the increased harassment, we next analyze the composition of harassers targeting the reincarnated VTuber. We categorize them based on their prior status with the original VTuber: \one Original Harasser (harassed the original VTuber), \two Original Viewer (were non-harassing viewers of the original), and New Viewer (no prior history with the original). Figure \ref{fig:harassment_viewer_composition} illustrates the composition of harassers for the 12 VTubers.

The results reveal that post-reincarnation harassment is primarily driven by individuals with a pre-existing connection to the VTuber.
First, a significant portion of the harassment comes from original harassers, who on average account for 37.2\% of the offending users. This confirms that some harassers are successful in tracking the Nakanohito across identities and are committed to perpetuating their harassment.
Second, a similarly large portion, averaging 41.4\%, consists of original viewers. This confirms that the contentious circumstances surrounding a reincarnation can convert formerly supportive fans into antagonists.
Finally,  new viewers constitute a small portion source of harassment, making up only 21.3\% on average. 

Overall, this compositional analysis reveals a dual threat for the reincarnated VTuber: they are not only pursued by legacy detractors but must also contend with a new wave of antagonism from a fractured segment of their once-loyal fanbase, creating a more complex and challenging environment to manage.

\begin{table*}
\centering

\footnotesize
\begin{tabular}{|L{9.5em}|l|L{36em}|L{16em}|}
\toprule
\textbf{Code Name} & \textbf{\%} & \textbf{Definition} & \textbf{Examples (translated)} \\
\midrule

Blaming for Abandonment & 10.7\% &
Messages that blame the Nakanohito for the discontinuation of their original VTuber persona, or accuse that they have betrayed or emotionally harmed the original fans. &
\textit{Don't run away, you coward.} \\ 

\midrule 

Accusations of Professional and Relational Damage & 5.8\% &
Messages that accuse the Nakanohito of causing damage to their former agency, the broader VTuber community, or specific relationships with former colleagues as a result of their reincarnation. &
\textit{(swearing), not going to have the collaboration with OOO} (a previous colleague VTuber) \textit{forever.} \\ 

\midrule

Attack on Past Controversies  & 8.3\% &
Messages that explicitly or implicitly bring up past controversies, scandals, or negative history to attack the performer. This includes those associated with the original VTuber persona or the Nakanohito. &
\textit{Holding a graduation concert just to farm for retirement money} (and then reincarnated) \textit{... that’s just typical of you} \\

\midrule

Attacks on the Current Viewership & 6.5\% &
Messages where the primary target of the insult or mockery is the current audience watching the reincarnated VTuber. &
\textit{I guess you can make millions just by tricking these simps} \\

\bottomrule
\end{tabular}
\caption{Codebook for inductive coding of harassment chat messages. The rest 68.8\% are those not relevant to reincarnation.}
\vspace{-6ex}
\label{tab:codebook}
\end{table*}

\vspace{-2ex}
\subsection{Harassment Content Analysis}
We now examine the specific content of the harassment chat messages after reincarna. By analyzing the themes within these messages, we can better understand the motivations behind the harassers' actions, \ie what is causing their anger or dissatisfaction.

\pb{Method.}
We employ a qualitative approach based on manual review. 
We construct a set of 1200 harassment chat messages by randomly sampling 100 for each reincarnated VTuber. The authors then manually evaluated these messages. 
The review process proceeds in two steps: First, each message is categorized to determine whether it is relevant to the reincarnation. Second, all messages deemed relevant to the reincarnation are subjected to an inductive coding process to identify recurring themes. 
We choose this method because many of these messages require sophisticated and extensive knowledge about the VTuber industry and community to interpret, and our preliminary experiments show that it is difficult for LLMs or topic models to capture the nuances.

\pb{Result.}
The coding yields four principal and distinct themes of harassment directly linked to the reincarnation event, as detailed in Table \ref{tab:codebook}. 
Overall, these four themes reveal the core motivations behind post-reincarnation harassment, with each category representing a different manifestation of the harasser's perceived grievance. 
These grievances are direct responses to the varied perceptions of the relationship between the virtual persona and the Nakanohito, and the different kinds of losses caused by the reincarnation.

The first two themes stem from a sense of direct harm: ``Blaming for Abandonment'' is a reaction to the emotional damage to the parasocial bond, while ``Accusations of Professional and Relational Damage'' is a response to the harm done to the Nakanohito's former brand and colleagues.
The third theme, ``Attack on Past Controversies'', exposes the perception that fuels this enduring hostility: the Nakanohito is the primary and continuous identity, no matter how the virtual persona changes. From this perspective, historical conflicts are not tied to a disposable VTuber persona but to the Nakanohito themselves, making them perpetually accountable for a history that reincarnation cannot erase.

Finally, ``Attacks on the Current Viewership'' may be understood as a direct result of the schism created within the original fanbase. The reincarnation forces a division: while some original fans feel betrayed, others who hold a different perception of the Nakanohito's actions successfully migrate and form the core of the new community. For those who feel abandoned, the existence of this loyal group can invalidate their own sense of loss.

%% file: sections/7.Conclusion.tex
\vspace{-2ex}
\section{Discussion \& Conclusion}
% \vspace{-0.5ex}
This paper provides a data-driven account of VTuber reincarnation, revealing it to be a high-stakes process fraught with significant challenges. We quantify the severe loss of audience and revenue for the Nakanohito (\S\ref{sec:rq1}), and the resulting net destruction of value for the agency and the VTuber industry (\S\ref{sec:RQ2}). Additionally, harassment can follow the Nakanohito and persist across different personas, and new forms of harassment may emerge because of the reincarnation (\S\ref{sec:rq3}). 
Together, these findings exemplify and highlight a core structural problem in the corporate VTuber model: the fraught relationship between the Nakanohito and the persona.
The Nakanohito's irreplaceability is the VTuber industry's Achilles' heel.
Our results suggest this core tension drives the instability, financial loss, and social friction of reincarnation events.
Below, we discuss how this central problem manifests in the business model and in audience perception, and conclude with recommendations for more effectively addressing reincarnation challenges.

\pb{Inherent Contradiction of the Corporate VTuber Model.}
This structural problem creates a fundamental contradiction within the corporate business model. The agency owns the IP (the virtual avatar), but the Nakanohito generates most of the value through their performance and parasocial bonds. Our results demonstrate the consequences of this schism: when the Nakanohito terminates, the value they created evaporates. It cannot be transferred to a new VTuber or effectively leveraged by the agency, nor can it be legally carried over to a new venture by the Nakanohito. This leads to the severe and irrecoverable losses of audience and revenue quantified in RQ1 and RQ2. The termination of a top talent does not merely redistribute assets within the market; it represents a net destruction of value, exposing the business model's profound fragility. In fact, consistent with our conclusion, the termination of Gawr Gura, who had the largest number of followers (4.7M), resulted in a 10\% fluctuation in her agency's stock price \cite{weekender2025gawr}.

\pb{Inconsistent Viewer Perceptions.}
This same ambiguity between the Nakanohito and virtual persona is mirrored in the audience, leading to inconsistent and often conflicting perceptions that drive community dynamics. These differing views coexist peacefully when the VTuber is operating normally. However, termination and reincarnation force viewers to realize their primary attachment: to the Nakanohito, the persona, or the combination. 
This divergence fractures the fan community. The Nakanohito-centric fans migrate, forming the bedrock of the new channel (\S\ref{sec:rq1}, \S\ref{sec:RQ2}). The persona-centric fans, whose object of affection has ceased to exist, largely disengage and exit the ecosystem (\S\ref{sec:RQ2}). Most critically, some fans see the reincarnation as a personal betrayal, converting their loss into targeted harassment (\S\ref{sec:rq3}). This schism is therefore a deep conflict rooted in  fundamentally different interpretations of the relationships of the VTuber, the Nakanohito, and viewers.

\pb{Recommendations.}
While this fundamental tension between virtual persona and performer may be unresolvable within the current corporate framework, our findings point to strategies that may partially mitigate the fallout for the stakeholders. For the reincarnated VTuber, success depends on a dual approach: first, carefully guiding their Nakanohito-centric core by using all permissible channels to subtly hint at their new identity, confirming the reincarnation as quickly as possible to close the information gap. This migration could be accelerated by platform recommendation algorithms capable of identifying and connecting displaced audiences. They also need robust platform support to combat the targeted harassment that follows them. For agencies, the clear imperative is talent retention: investing in talent satisfaction, fair contracts, and Nakanohito wellbeing. On the other hand, when termination and reincarnation occur, they must rely on the intra-agency ecosystem \ie the other VTubers on their roster, to retain displaced fans. Similarly, for other VTubers, these events represent a crucial opportunity to attract a high-value, unattached audience by being visible and actively welcoming. For both groups, their ability to capture this ``whale fall'' audience could be significantly enhanced by recommendation systems designed to surface relevant alternative content to these high-value, unattached viewers. 

\pb{Limitations \& Future Work.}
We acknowledge several limitations that offer avenues for future research. First, our analysis focuses on the direct before-and-after effects for the reincarnated VTuber; future work could contextualize these findings by comparing them against a global baseline of industry-wide viewership trends. This would help isolate the specific impact of reincarnation from broader market fluctuations, such as post-pandemic shifts in online entertainment consumption. Second, our sample of 12 cases is intentionally focused on high-profile performers from major agencies. To enhance generalizability, subsequent research should expand this scope to include independent VTubers and talents from smaller corporations, exploring how these dynamics affect the average digital laborer. Finally, our operationalization of viewer migration could be refined. Future studies might develop more nuanced models that account for common community behaviors, such as sporadic, event-driven financial support --- as opposed to consistent subscriptions --- and the use of alternate user accounts to follow a new persona, which could provide a more precise measure of true fan retention.

%% file: sections/A.Appendix.tex
\clearpage
\appendix
\section{Appendix}

\subsection{Ethical Considerations}

\pb{Data Collection and User Privacy.}
We collected data from publicly accessible YouTube sources. Our collection process was designed to be responsible and non-disruptive; data requests were rate-limited and distributed over a four-month period to avoid imposing an undue load on the platform's servers. The collection was entirely observational and did not involve any direct interaction with viewers or VTubers. All user-identifying information was immediately anonymized using a non-reversible hashing algorithm, and our findings regarding viewers are presented in aggregate to ensure individual privacy is protected. Our practice also aligns with previous research using the same data source \cite{chi25who_reaps}.

\pb{Anonymity of the Reincarnated VTubers.}
A central ethical decision made for this study is protecting the identities of the performers (Nakanohito). We have deliberately avoided any language or data presentation that would explicitly link an original VTuber persona to its reincarnated form. While these connections may be considered ``common knowledge'' within fan communities, our research respects industry conventions and potential non-disclosure agreements by not serving as a source of confirmation. 
All VTubers in our 12 case studies have been anonymized (\eg V1, R1). The identities of the terminated VTubers can be discerned, as they are public figures and their termination is officially announced. However, we ensure that we do not disclose information that could link to their reincarnated identity.

\subsection{Data Description}

\begin{table}[h!]
\small
    \centering
    \begin{tabular}{|llL{18em}|}
    \toprule
        \textbf{Filed} & \textbf{Type} & \textbf{Description} \\
    \midrule
        ID & String & A unique identifier for the livestream session \\
    \midrule
        Title & String & Plain text title of the livestream session. \\
    \midrule
        Description & String & Plain text description of the livestream session. \\
    \midrule
        Duration & Integer & The total duration of the livestream session, measured in seconds. \\
    \midrule
        Release & Float & A Unix timestamp for the start time of the livestream session. \\
    \midrule
        Channel & String & The unique identifier of the YouTube channel (also uniquely identifies the VTuber) where the livestream session happens  \\
    \bottomrule
    \end{tabular}
    \caption{Description of data field for livestream session records.}
    \label{tab:live_description}
\end{table}

\begin{table}[h!]
\small
    \centering
    \begin{tabular}{|L{4em}lL{18em}|}
    \toprule
        \textbf{Filed} & \textbf{Type} & \textbf{Description} \\
    \midrule
        ID & String & The unique identifier of the interaction record \\
    \midrule
        LID & String & The unique identifier of the livestream session where the interaction happens \\
    \midrule
        UID & String & The unique identifier of the user of the interaction \\
    \midrule
        Kind & Enum & Chat, Superchat, membership, or gift membership \\
    \midrule
        Offset & Integer & When the interaction happens in the livestream session, measured in second \\
    \midrule
        Money & String & The paid amount and the currency for the interaction. \\
    \midrule
        Member Status & String & Membership status of the user of the interaction \\
    \bottomrule
    \end{tabular}
    \caption{Description of data field for viewer interaction records.}
    \label{tab:interaction_description}
\end{table}

\newpage